\documentclass[sigconf]{acmart}

\newcommand{\learners}{\textbf{\textit{Learners}}}
\newcommand{\novice}{\textbf{\textit{Novice}}}
\newcommand{\novices}{\textbf{\textit{Novices}}}

\newcommand{\CtrlC}{$\mathrm{Control}_C$}
\newcommand{\CtrlS}{$\mathrm{Control}_S$}

\definecolor{avtgreen}{RGB}{0, 150, 85}

\usepackage{subcaption}

\AtBeginDocument{%
  \providecommand\BibTeX{{%
    \normalfont B\kern-0.5em{\scshape i\kern-0.25em b}\kern-0.8em\TeX}}}

\copyrightyear{2024}
\acmYear{2024}
\setcopyright{acmlicensed}\acmConference[SIGCSE 2024]{Proceedings of the 55th ACM Technical Symposium on Computer Science Education V. 1}{March 20--23, 2024}{Portland, OR, USA}
\acmBooktitle{Proceedings of the 55th ACM Technical Symposium on Computer Science Education V. 1 (SIGCSE 2024), March 20--23, 2024, Portland, OR, USA}
\acmPrice{15.00}
\acmDOI{10.1145/3626252.3630887}
\acmISBN{979-8-4007-0423-9/24/03}

\begin{document}

\title{Learners Teaching Novices: An Uplifting Alternative  Assessment}

\author{Ali Malik}
\email{malikali@cs.stanford.edu}
\orcid{0009-0007-1201-5014}
\affiliation{%
  \institution{Stanford University}
  \state{California}
  \country{USA}
}

\author{Juliette Woodrow}
\email{jwoodrow@stanford.edu}
\orcid{0009-0006-8097-093X}
\affiliation{%
  \institution{Stanford University}
  \state{California}
  \country{USA}
}

\author{Chris Piech}
\email{piech@cs.stanford.edu}
\orcid{0000-0001-5140-0467}
\affiliation{%
  \institution{Stanford University}
  \state{California}
  \country{USA}
}

\renewcommand{\shortauthors}{Ali Malik, Juliette Woodrow, \& Chris Piech}

\begin{abstract}
We propose and carry-out a novel method of formative assessment called Assessment via Teaching (AVT), in which learners demonstrate their understanding of CS1 topics by tutoring more novice students. AVT has powerful benefits over traditional forms of assessment: it is centered around service to others and is highly rewarding for the learners who teach. Moreover, teaching greatly improves the learners' own understanding of the material and has a huge positive impact on novices, who receive free 1:1 tutoring.  Lastly, this form of assessment is naturally difficult to cheat---a critical property for assessments in the era of large-language models.
We use AVT in a randomised control trial with learners in a CS1 course at an  R1 university. The learners provide tutoring sessions to more novice students taking a lagged online version of the same course. We show that learners who do an AVT session before the course exam performed 20 to 30 percentage points better than the class average on several questions. Moreover, compared to students who did a practice exam, the AVT learners enjoyed their experience more and were twice as likely to study for their teaching session. We believe AVT is a scalable and uplifting method for formative assessment that could one day replace traditional exams.

\end{abstract}

\ccsdesc[500]{Social and professional topics~Computing education}
\ccsdesc[500]{Social and professional topics~Student assessment}

\keywords{Formative assessment,
student-led teaching,
peer teaching,
studying strategies,
learning at scale,
online courses
}

\maketitle

\section{Introduction}

Assessment, both formative and summative, is a cornerstone of education. However it is one of the most frustrating parts of learning for both students and teachers. Students are forced to partake in experiences -- such as written or oral exams -- that are stressful, adversarial, and may not feel meaningful. At the same time, teachers have to spend hours creating and grading exams, all while contending with the ever growing possibility of cheating: a problem of pressing importance with the rise of Large Language Models.  All this together makes assessment a focal point of distrust and tension in the learning experience. 

Perhaps there is a better way. Consider the common insight that \emph{the best way to learn is to teach} and the parallel idea that \emph{you really understand the material when you can teach it}. These insights inspire an alternative assessment which is   authentic, oriented towards service, improves learner understanding, and is hard to cheat.

In this paper, we propose and carry-out a novel method of assessment called Assessment via Teaching (AVT), in which students ({\learners}) demonstrate and refine their understanding of CS1 topics by tutoring more novice students ({\novices}).  This approach is characterised by several key features:

\begin{itemize}
    \item Authentic and service-oriented, resulting in a fulfilling experience for {\learners}. 
    \item Provides rich, memorable feedback to {\learners} and deepens their understanding. 
    \item Can be repeatedly taken by {\learners} with no extra effort from course staff.
    \item Impactful for {\novices}, who get free 1:1 tutoring.
    \item Difficult for {\learners} to cheat or ``study towards the test''. 
\end{itemize}

\begin{figure*}
    \centering
    \includegraphics[width=0.95\textwidth]{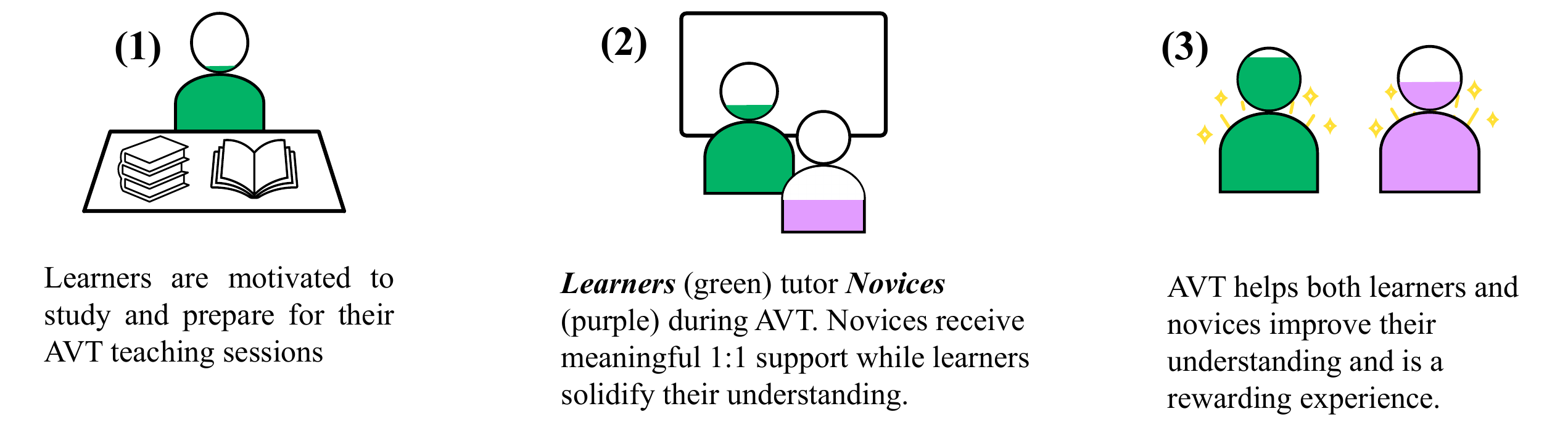}
    \caption{An overview of Assessment via Teaching (AVT). Colour fill of each character indicates understanding of material.
    }
    \label{fig:overview}
\end{figure*}

We carry out the first ever attempt at using AVT with real students through a randomised control trial (RCT). Specifically, we recruited learners in a CS1 course at a large R1 university and randomly assigned a subset to provide 1:1 AVT tutoring sessions to novice students. The novices were students from around the world taking a lagged, online version of the same course \cite{Data2023codeinplace}. We assigned the rest of the participants randomly into two control groups: (1) a set of people who were told to sit a 1-hour practice exam, and (2) a group of people who were never scheduled to do any session (practice exam or AVT). The former let us compare AVT to traditional studying strategies while the latter helped us understand the selection bias of our experimental population.

Our results quantitatively verify the benefits of AVT, showing that learners who teach an AVT session prior to an exam perform up to 30 percentage points higher than the class average \textit{on several questions}. Moreover, AVT learners signficantly enjoyed their experience more (effect size of +0.6 on a 5 point scale) and were more than twice as likely study for their session than those who did a practice exam. We also demonstrate the usefulness of the AVT sessions to the novices being helped and discuss how the AVT teaching interactions provide a useful signal of learner understanding to course instructors. We close with a detailed discussion about challenges, limitations, and future work.

The main contributions of this paper are as follows. In Section \ref{sec:avt} we propose Assessment via Teaching and discuss details for replication. In Section \ref{sec:exp} we report the quantitative and qualitative results of our randomised control trial.

Finally, in Section \ref{sec:challenges} and \ref{sec:discussion}, we discuss the challenges, limitations, and next steps of AVT.

\section{Related work}\label{sec:related} 

\paragraph{\textbf{Formative assessment}} 
Formative assessment, which emphasises helping students improve their understanding, has huge benefits for learning \cite{BlackWilliam98, ZEIDNER2007165, Grover21}. 

Many different formative assessments have been explored over the years, especially in the context of CS1 (see \cite{Grover21} for an overview).  However, to the best of our knowledge, our proposal is the first to use real teaching as an explicit approach to formative assessment.

\paragraph{\textbf{Student-led teaching}}
There is extensive research on the benefits of having learners take on a teaching role, including improved learner understanding (\cite{Bargh1980OnTC, Benware1984, Nestojko2014, Bruner72, gartner1971children}), higher motivation, and impact on socio-psychological factors like identity and belonging \cite{Goldschmid1976,Newcomb1967}. There is also a notable economic consideration: as class sizes grow larger, leveraging learners to teach can be a crucial step to scaling support for students \cite{Goldschmid1976, UndergradTA17, piech2021code}. This has been particularly important in CS1 education where enrolments have skyrocketed.

\paragraph{\textbf{Impact of 1:1 help to novice students}}
A key reason why AVT is meaningful for the learners who teach is its real impact on novice students. 
Receiving high-touch, 1:1, personalised help has been shown to be one of the most effective educational interventions a student can receive \cite{Bloom82TwoSigma}. Crucially, this help doesn't have to come from a professor. Several studies, in CS1 and more broadly, have shown that near-pear tutoring can be just as effective, if not more effective, than help from an expert \cite{maas1973students, Keller74, Reges03, Kulkarni15, piech2020co, piech2021code, reges1988effective,UndergradTA17,mirza2019undergraduate,roberts1995using}. In fact, receiving help from a learner who is more similar in age, demographics, and is more relatable to novices, has a profound impact on novice identity forming and belonging with the subject \cite{Bligh1972, Zander1974, Newcomb1967, UndergradTA17}.

\section{Assessment via Teaching (AVT)}\label{sec:avt}

\vspace{2mm}
\begin{quote}
    ``\emph{If you can't explain it simply, you don't understand it well enough.}'' -- Unknown
\end{quote}
\vspace{1mm}

It is common folklore in the teaching community that being able to explain a concept is a strong signal of understanding. 
How can we instantiate this intuition into an actual mechanism of formative assessment for learners? In this section,
we present Assessment via Teaching, which is comprised of two groups of people:

\begin{itemize}
    \item {\learners}: These are students of the course who we wish to formatively assess. They will be doing the teaching.

    \item {\novices}: These are novice students who are also learning the material of the course, but are several weeks behind the learners. They receive help from the learners.
\end{itemize}

Examples of novices could be students taking an online, lagged version of the course or peers who have fallen behind. We discuss the challenges of finding a novice population more in \S \ref{ssec:finding_novices}.

Each AVT assignment requires the learner to hold a 1-hour tutoring ``help session'' for the novices. The learners are told to prepare for a subset of topics e.g. control flow and function decomposition in a CS1 course. The learner then hosts their help session, wherein they are paired 1:1 with a novice student who is stuck on a problem related to these topics. During this session, the learner's job is to guide the novice towards solving their problem.  After a 1:1 session is concluded, the learner can be paired with a brand new novice, until the entire hour has elapsed. 
The key idea behind AVT as a formative assessment is this: \textbf{in order to support novices effectively, learners have to develop a strong understanding of the course material}. And in the process of helping the novices, the learners deepen their existing understanding and gain insight on concepts they need to continue studying.

\subsection{Benefits of AVT}
We present why we believe AVT is a powerful and uplifting  formative assessment. In Section \ref{sec:exp}, we verify many of these claims.

\paragraph{ \textbf {Fulfilling and motivating for learners }}

Assessment often harbours negative, adversarial feelings from students, who can feel like they are being judged and measured.

Compared to this, we believe AVT is more motivating for learners. This is because it combines \textbf{assessment} and \textbf{public service}, drawing on the learning benefits of former with the fulfilment and motivation of the latter.
Since the end goal is to help a real novice student, learners feel a strong sense of responsibility to study the material in preparation. This is deepened by a stake of personal identity and embarrassment of letting down their novice students. Moreover, a successful AVT session has an observable impact on novices---a feeling that is gratifying for the learner hosting the session.

\paragraph{ \textbf {Naturally formative feedback }}
Having to explain a concept is a powerful method to deepen understanding \cite{Bargh1980OnTC}. In AVT, explaining concepts effectively to a novice requires a thorough understanding of the course material. Novice students often have fundamental misunderstandings that require teachers to have a strong mental model of the concepts. As learners engage with novice questions, they refine their own mental model and, in the process, uncover new misunderstandings of their own. In addition, when mistakes in understanding are unearthed and resolved in teaching, these moments are extremely memorable and leave a lasting impression in the teacher's mind.

\paragraph{\textbf{Repeatable test}} 

Allowing students to repeatedly retake assessments is extremely desirable \cite{Grover21}. However, the practicality of generating novel questions or activities for every instance of retesting is a huge challenge. AVT resolves this predicament by providing a standardised yet dynamic evaluation platform that accommodates repeated assessments without necessitating the continuous creation of new tests. In theory, learners should be able to redo AVT as much as they want, resulting in a constant improvement of their understanding while helping more novices.

\paragraph{ \textbf {Impactful 1:1 help for novices}}
High-touch, personalised 1:1 help for students is one of the most impactful educational interventions known to researchers \cite{Bloom82TwoSigma}. Unfortunately, 1:1 tutoring is prohibitively expensive for most students.
Multiple investigations have shown that early learners can be incredibly effective tutors if given the right support \cite{maas1973students, Keller74, Reges03}.  With AVT, we channel the efforts of bright students, who would otherwise waste hours sitting an exam, into instead providing a meaningful service of tutoring novices. Such a system, if implemented broadly, could support 1:1 help for learners at a global scale, especially those with limited access to resources.

\paragraph{ \textbf {Hard to over-optimise or cheat }}
Assessment often suffers from the famous Godhart's law: ``when a measure becomes a target, it ceases to be a good measure''. For many assessments, 

students begin to optimise by studying for the test over improving their understanding. 
One benefit of AVT is that there is no clear way to study for the assessment itself. If a student just wishes to perform well at AVT, their best bet is to actually improve their understanding of the material. 

Cheating is another major concern in assessment, especially in the new era of large-language models (LLMs) \cite{Milano2023}. Several assessment platforms have started adding biometric checks like video recording to ensure an individual is completing an assessment by themself. There is also an increasing distrust of written exams in general, with assessors instead opting for some sort of oral interview where a participant's understanding can be adaptively probed. This is effective but impossible to scalable.
AVT naturally integrates both these dynamics into its design. AVT is a live help session, which automatically makes it a biometric verification, and helping novices with their issues is similar to an adaptive oral interview where the learner must draw on their understanding dynamically in response to issues as they arise.

\section{Case Study: AVT in a CS1 Course}\label{sec:exp}

To better understand the potential of AVT, we ran a randomised control trial (RCT) with learners in a CS1 course at a large R1 university. In this course, students learn about the fundamentals of programming, such as control flow, functions, console, graphics, and animation. The course had an enrolment of about 300 students and ran for ten weeks.
Alongside the university course, there was an online version of the exact same class with around 10,000 students from across the globe. This online version also happened to be three weeks behind the university course, presenting a perfect source of {novices} for the learners to teach.

\subsection{Experimental Setup} Approximately two weeks before the university course's midterm and final exam, we advertised a learning experiment where students could enroll in a study on "learning and studying strategies" for CS1. To control for selection bias, we did not reveal details about the experiment. 

After a week long enrollment period, we took the list of eligible participants and randomly selected a subset to be the {\learners} who teach in our AVT experiment. The remainder of the students were randomly assigned to one of two control groups (discussed below).
The {\learners} were asked to prepare for an hour long AVT session where they would be tutoring \textit{real} novice students on a problem related to specific topics\footnote{The {learners} were aware that the {\novices} were authentic learners participating in a free online course}. For the midterm exam, the topics were control flow, function decomposition, and pre/post conditions. For the final exam, the topics were graphics, animation, and keyboard interactions. The {learners} were not told ahead of time the specific question they would be helping with.

The {\novices} in our experiment were sourced from an online version of the university course with about 10,000 students. This online course had the ability to offer just-in-time help to students, which we used to schedule novices for each AVT session. 

At the time of the AVT session, the learner would host a video conferencing call and would be joined by a novice student working on a particular question. {\learners} had the ability to see the {\novice} student's source code, but could not directly make edits \cite{tj24_pyodidu}. To control for variance, all the AVT sessions for the midterm involved novices working on the exact same assignment, and same for the final. The learner would then help the novice with whatever they were stuck on. If the problem was solved quickly, another novice would join until the entire hour was complete. All the AVT sessions were recorded for safety and analysis purposes.

\paragraph{\textbf{Outcome variables}}
To understand the benefits of AVT, we collected at a few outcome variables. Firstly, for each {learner}, we looked at their eventual score on the midterm/final exam on the question relevant to the topics of their AVT sessions. For the midterm, this was Q1, and for the final this was Q3. We also sent the {learners} an exit survey to collect information about how they prepared for their AVT sessions and how enjoyable they were. Lastly, expert instructors manually analysed the AVT session recordings to quantify how much understanding learners demonstrated, how effective they were at teaching, and whether the sessions were helpful for the novices.

\begin{table}
    \centering
    \begin{tabular}{lll}
        \toprule
        \textbf{Group name} &  \textbf{Description} & $n$ \\ 
        \midrule 
        \CtrlC  &  Regular students in the university course. & $244$ \\
        \CtrlS   &  Enrolled in experiment but did no session. & $17$ \\
        PracticeExam &  Did a practice exam session. & $14$\\
        AVT &  Taught an AVT session.  & $15$\\
        \bottomrule
    \end{tabular}
    \caption{Experimental groups for our analysis. }
    \label{tab:expgroups}
\end{table}

\paragraph{\textbf{Control groups}}

To compare the AVT sessions to a strong baseline of studying methods for exams, we had a random subset of participants sit a 1 hour practice exam session instead of doing AVT. In this practice exam, they were given the same question discussed in the AVT sessions. This group will be referred to as \textit{PracticeExam}.
We also wanted to control for the selection bias of people willing to join an experiment, so we chose a random subset of participants to not do any kind of session (AVT or practice exam).  This group will be called the \CtrlS{} group. 
Lastly we had the \CtrlC{} group, consisting of students who did not interact with our experiment at all. 
The summary of these groups can be seen in Table \ref{tab:expgroups}.

\begin{table}
  \centering
  \begin{tabular}{lcccccc}
  \toprule
  &  \multicolumn{6}{c}{\textbf{Question Scores (percentage points)}}  \\
  \cmidrule{2-7}
  & Q1 & Q2 & Q3 & Q4 & Q5 & Q6 \\
  \midrule

\textbf{{Midterm}} &&&&&& \\

 \CtrlC  & 64.9 & 75.5 & 64.7 & 61.6 & 58.0 & - \\

 AVT & {\color{avtgreen} \textbf{89.3}} & 89.3 & \textbf{94.0} & \textbf{79.3} & 72.5 & - \\
 $p$-value & ({\color{avtgreen}0.004}) & ({0.127}) & ({0.000}) & ({0.016}) & ({0.078}) & - \\ 

\cmidrule{1-7}
\textbf{{Final}} &&&&&& \\
  \CtrlC & 79.4 & 70.4 & 81.4 & 89.0 & 75.6 & 76.0 \\
  AVT & 82.1 & 84.6 & {\color{avtgreen}\textbf{96.2}} & \textbf{98.3} & \textbf{92.9} & \textbf{89.5} \\
  $p$-value & ({0.292}) & ({0.051}) & (\color{avtgreen}0.000) & (0.000) & ({0.001}) & ({0.004}) \\  
\bottomrule
\end{tabular}
\caption{Teaching an AVT session significantly improves exam performance on more than half the questions. Scores given as percentage points out of 100 and bolded for $p < 0.05$. Green indicates questions related to the AVT topics.}
\label{tab:exam_perf}
\end{table}

\begin{figure}[t]
  \centering
  \includegraphics[width=0.95\linewidth]{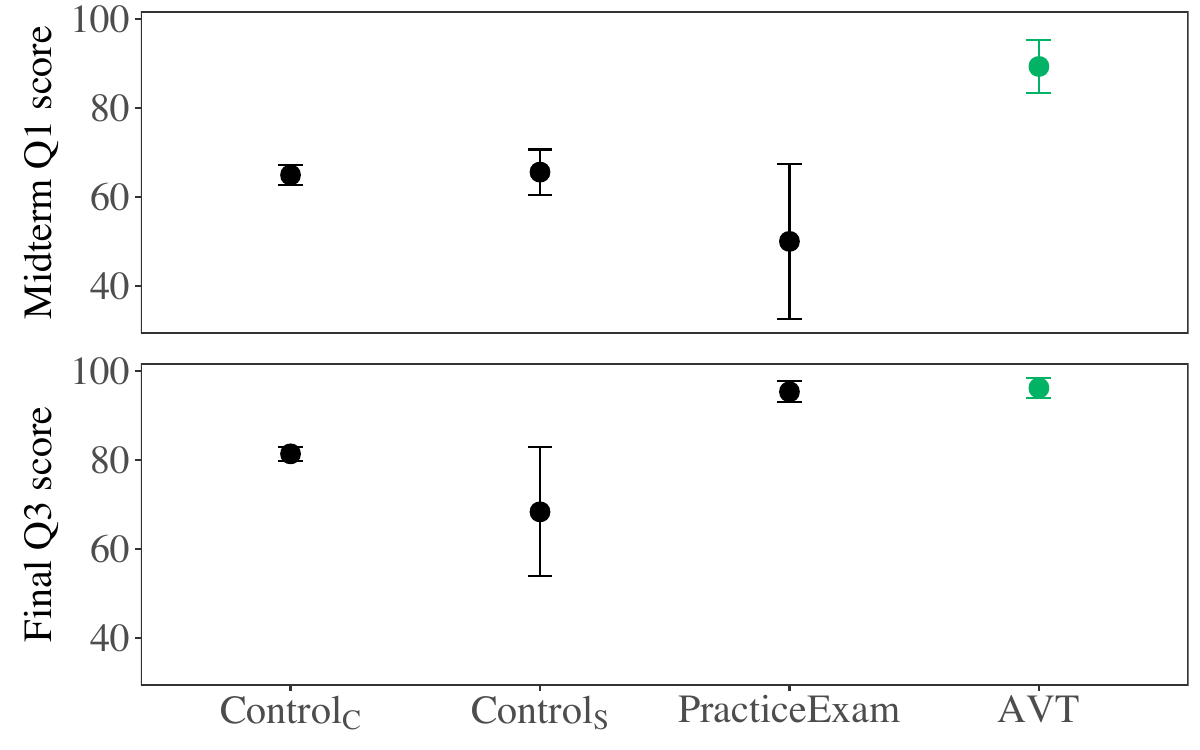}
  \caption{The AVT group preforms better on the relevant exam question compared to class average and as good as the practice exam group. This isn't due to a selection bias between the course (\CtrlC{}) and experiment (\CtrlS{}) populations.}
  \label{fig:protege}
\end{figure}

\begin{figure}
  \centering
  \includegraphics[width=\linewidth]{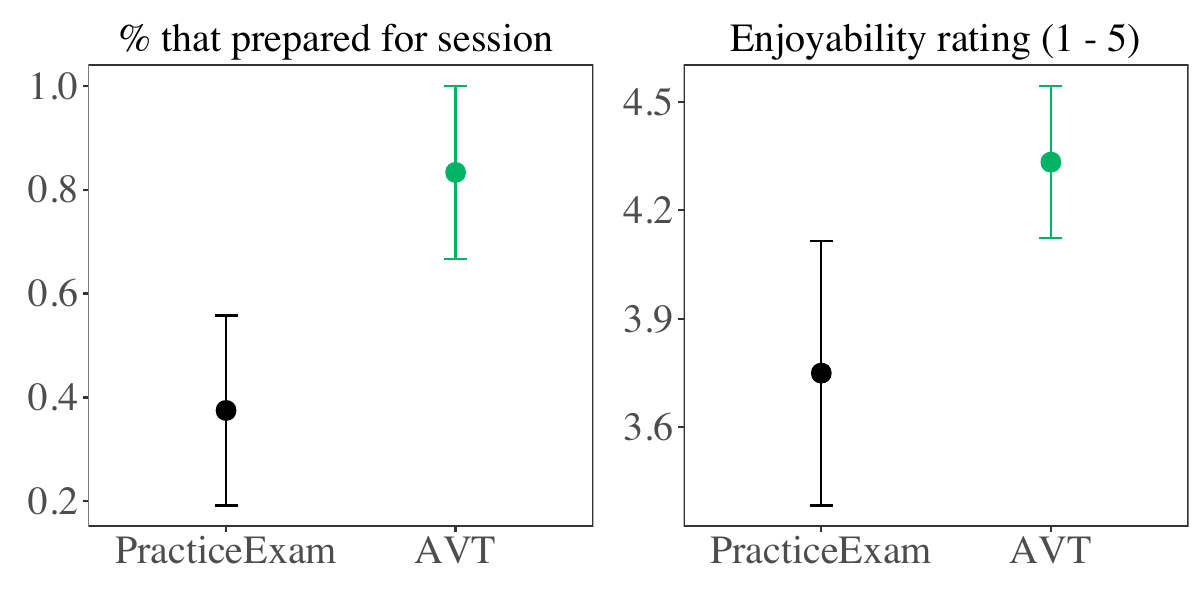}
  \caption{Learners are both more motivated to prepare for AVT sessions and enjoy the sessions more compared to doing a traditional practice exam session.}
  \label{fig:survey}
\end{figure}

\subsection{Analysis}\label{ssec:analysis}

We analysed the results from our experiment to better understand the value of AVT as a formative assessment for {\learners} and {\novices}:

\paragraph{ \textbf{Learners: Beneficial for understanding}}

To investigate the helpfulness of AVT in improving learner understanding, we looked at the performance of learners in the midterm/final exam following their session.
Table \ref{tab:exam_perf} shows the average score (in percentage points) obtained by the AVT learners on each question in the midterm and final exam, as well as the class averages (\CtrlC). 

We compute a one-sided $t$-test for the significance between the AVT score and the class average, with $p$-values provided in parentheses.\footnote{We compare to this group due to larger sample size but our results hold even when compared to \CtrlS{} as well.}

Our results suggest a strong effect of AVT teaching on learner understanding. In 3 out of 5 questions on the midterm, the AVT group outperformed the class average ($p <0.05$), with improvements of 25 to 30 percentage points (p.p.). A similar trend is seen in the final, although the improvements are in the 10-15 p.p. range\footnote{Note: The final exam was much easier than the midterm, with higher average score and lower variance in scores}. Reassuringly, we always see a significant increase in  questions directly related to the AVT topics (Midterm Q1 and Final Q3).

Next we looked at whether these effects were due to a selection bias in our experiment population. In Figure \ref{fig:protege}, we compare performance on Midterm Q1 and Final Q3 across all our experimental groups. We see that there is no significant difference between the class average and our experimental participants (\CtrlC{} vs \CtrlS{}).
We can see also see that AVT was \textit{at least} as effective as doing a practice exam---an extremely compelling outcome, since practice exams are known to be good ways to prepare \cite{Balch98}.\footnote{We note that in the midterm, the PracticeExam group seemed to preform badly, but we believe this is a statistical anomaly due to small sample size ($\approx$ 4 students).}

\paragraph{ \textbf{Learners: AVT is motivating and enjoyable.}} 
We also explored how motivated the {learners} were to study for their AVT sessions. To do this, our exit survey asked participants how long they spent preparing for their session. Figure \ref{fig:survey} (left) shows the fraction of participants in the AVT group who spent more than 1 hour preparing for the session compared to those in the PracticeExam group. We can see this fraction is almost twice as high as for the AVT group, suggesting that AVT participants took their preparation more seriously.
We also tried to measure how enjoyable AVT was for learners. Figure \ref{fig:survey} (right) shows that leaners found the AVT sessions more enjoyable compared to those who did a practice exam. In addition to this, the AVT group also gave an average recommender rating of 4.2 out of 5 for their experience and provided some positive qualitative feedback:

\begin{quote}
\textit{
    ``I had a lovely experience! I was grateful that it forced me to prep for graphics/animation for the final, and my student was so nice :). I felt encouraged in my teaching skills and programming knowledge.''
    \\
    --- Learner 1
    \\
}
\end{quote}

\begin{quote}
\textit{
    ``Having to explain graphics/animations concepts to another person helped me realize my own weaknesses.'' \\
    --- Learner 2
}    
\end{quote}

\paragraph{ \textbf{Novices: AVT help is impactful}}

Expert instructors manually assessed each AVT session to measure the impact on novices that the AVT learner made. The experts checked for two things: (1) Did the novice make significant progress on their problem?  (2) Was it because the learner just gave away the answer?

In all but one session (98\%), the experts agreed that learners helped the novices make progress and did so in an educationally helpful way. For the one session where this was not the case, the novice student was particularly unprepared and struggled with basic syntax.
That being said, the experts did notice signs of some learners being more comfortable ``telling'' rather than ``guiding''.  This is a major concern if AVT is to be used more broadly, and we discuss it further in \S \ref{ssec:untrained_teachers}.

\section{Implementation challenges}\label{sec:challenges}

While AVT seems promising as a meaningful and fulfilling formative assessment, it is a novel proposal and several challenges need to be addressed if it is to see widespread usage in diverse contexts.

\subsection{Finding a population of novices}\label{ssec:finding_novices}

A crucial aspect of AVT is it's authentic, service-oriented impact on real novice students. For this to work, we need access to a population of novices who the learners can teach. Ideally these novices are (1) reliably available, (2) appropriately levelled so that they need help with content most relevant to learners, and (3) verified, so that the interactions are safe.

Where can we find novices if we are to use AVT in more CS1 contexts?  While there isn't an existing out-of-the-box solution, we believe the true number of novices that exist in the world who would be interested in 1:1 help is substantial. Some insightful data points come from Philip Guo's Python Tutor: an online platform where you can ask for Python help and be matched with a human being in seconds, which has been used by over 10 million people \cite{GuoPythonTutor13, GuoPythonTenYears21}. A related platform is Schoolhouse.world, a sister company of Khan Academy that provides free, online tutoring to students around the world. Schoolhouse has around 48k students. 

Another avenue for reaching novices would be existing MOOC platforms, which have a consistent stream of students learning new courses. This year alone saw a record high of 100 million registered users on Coursera. Lastly, there are also opportunities for tutoring novices in understaffed or low-resources schools around the world or students in the university who have fallen behind on the course.

\subsection{Importance of domain fluency in CS1}

Having learners do AVT sessions could be an uncomfortable experience for people who are more introverted or feel reluctant to take on a position of authority with novices.  
For learners, having a difficult experience could be particularly damaging to their identity, especially when starting out in a new subject like CS1.

One response to this is the a belief that beyond the course material, the ability to teach and fluently communicate CS1 concepts is an important skill for learners to develop. In this sense, AVT is both formative for content understanding and also for leadership and communication skills. The other approach is to offer AVT as just one option out of many possible kinds of formative assessment in a course. We believe many learners would find the service-oriented impact of AVT extremely compelling.

\subsection{Safety}
Safety and mental security of the learners doing AVT is of the utmost importance, especially for learners early in their identity-forming phase of a new subject. An unpleasant teaching experience or a toxic interaction could have detrimental downstream effects on the learners in charge. This is of particular concern if the pool of novices is unvetted, such as in an open, online setting.

In our experiment, we monitored the interactions during the session to audit them for any inappropriate behaviour. This safety measure can easily be automated using cloud recordings and LLM based evaluation of the transcripts for negative behaviour. Moreover, we had a reporting feature for our AVT sessions that both novices and learners could use.

\section{Discussion}\label{sec:discussion}
In this section, we discuss some further ideas on AVT and what role it could play in the future of assessment.

\subsection{Altruistic and uplifting assessment}

Assessment stands as the central gatekeeping mechanism within education, wielding considerable influence over the learning process. Traditional examination formats, although effective for certain students, may not resonate as effectively with individuals motivated by a desire to contribute to the welfare of others. AVT instantiates assessment through a  service-oriented, altruistic pursuit. The focus shifts from individual evaluation of the self to the betterment of one's understanding in order to support others. This has the potential to reshape students' perceptions of the purpose of computer science education. Our hypothesis is that the broader societal emphasis of AVT could serve as an effective tool for fostering inclusivity among traditionally underrepresented groups.

\subsection{Potential as a summative assessment}\label{ssec:summative}

Could AVT go one step further and replace traditional summative exams? While the answer is not clear yet, we believe this is a promising research direction.

\paragraph{\textbf{One-sided signal of understanding}}

The first indication of this possibility is a small investigation we conducted in which two expert teachers watched the midterm AVT interactions and tried to predict what kinds of mistakes the learner would or would not have made in the exam. A validation showed that the expert predictions were correct in practically all cases. However, there was a one-sidedness to their correctness: the experts could easily predict when a learner understood something well and likely wouldn't make a mistake, but it was much harder to make strong claims about what the learner \textit{didn't} know.

This asymmetry is a notable difference from traditional exams, where the set of questions help instructors see what students both do and don't understand.
This suggests that AVT could be used to ``grade-up'' rather than ``grade-down'', which has positive psychological benefits for learners. However, to get a comprehensive picture of the learner understanding would need the support of additional assessment or a different scheme for AVT.

\paragraph{\textbf{Quantifying understanding in AVT}}
It is also a challenge to find a reliable, affordable, and unbiased way to measure learner understanding from AVT sessions. In our experiment, we had expert instructors review the sessions, which is unsustainable at scale.
One idea is to try grading learners based on how much their novice students improved after the AVT session. This would require assessing the novices in a more traditional way, through an autograded question before and after the AVT session. 
Another solution is to leverage the power of large-language models (LLMs) to go through the AVT session transcripts and identify moments of demonstrated learner understanding. This would require in-depth research to ensure the outputs were meaningful and fair for learners.

\paragraph{\textbf{Stochasticity of the novices}}

If we are to assess learners summatively using AVT, we need to account for the stochastic variance of the novices they are paired with; it might be easier to tutor someone who already has a good grasp of the material compared to a novice who is extremely confused. To this end, one could assess a learner  over \textit{several} AVT sessions with multiple novices to reduce variance. Alternatively, rather that randomly pairing novices with learners, we can use clever matching strategies to ensure the learner-novice pairing is ``good'' in some rigorous sense.

\subsection{ \textbf{How good are untrained people at teaching?}}\label{ssec:untrained_teachers}
A key theme in several papers in CS1 education has been the remarkable ability of young undergraduates to be effective teaching assistants \cite{UndergradTA17}. We build on this direction by trying to understand how effective a CS1 student, who has just learned a new concept, is at supporting other more novice students. This question sits directly between the literature on peer teaching and undegraduate TAs.
We manually evaluated the teaching interactions of our AVT sessions based on how well we thought the learner did on guiding the novices. As mentioned in \S \ref{ssec:analysis}, the teaching interactions had a largely positive impact on the novices, showing how effective even untrained students can be as a source of support for novices.
That being said, we found some instances where the teaching interactions could have been improved. Several learners would try to recite the answer to novices rather than guide them to figure out the problem on their own. We observed that learners were very hesitant when novice code was buggy, almost as if they felt personally judged the code not working.
These findings suggest that while AVT has positive benefit for novices, it would be beneficial to provide learners with some form of teacher training.

\section{Conclusion}
In this paper, we presented a new alternative to traditional assessment that is simultaneously fulfilling, impactful, repeatable, hard to cheat, and highly effective in improving learner understanding.  We showed strong quantitative evidence for these benefits and discussed at length the challenges to be solved before this approach can see widespread usage. While there is still some way to go, we truly believe that AVT is a scalable and uplifting method for formative assessment that could one day replace traditional exams.

\begin{acks}
Ali is supported by a graduate fellowship award from Knight-Hennessy Scholars at Stanford University.

We thank the following people for their help on different parts of the project: Chubing Li, Aya Mouallem, Julia Markel, and TJ Jefferson.
\end{acks}

\bibliographystyle{ACM-Reference-Format}
\bibliography{ref}

\end{document}